\documentclass[10pt,twoside]{amsart}
\usepackage{amssymb, amsmath, amsfonts}          
\usepackage{epsfig, url} 
\usepackage{amsmath, amssymb, amsthm}   
\usepackage{boxedminipage}
\usepackage{epsfig}
\usepackage{pstricks,pst-node,pst-plot}
\usepackage{setspace}

\newtheorem{theorem}{Theorem}[section]

\newtheorem{claim}[theorem]{Claim}
\newtheorem{lemma}[theorem]{Lemma}
\newtheorem{corollary}[theorem]{Corollary}

\newtheorem{question}[theorem]{Question}

\newcommand{\LCA}{\mathop{\rm LCA}}
\newcommand{\NDesc}{\#desc}
\newcommand{\Desc}{desc}

\newcommand{\Chain}{\mathop{\rm Chain}}
\newcommand{\tree}{\ensuremath{\mathcal{ T }}}

\newcommand{\cyc}{\ensuremath{\mathcal{ C }}}
\newcommand{\G}{\ensuremath{\mathcal{ G }}}

\newcommand{\comment}[1]{}

\def\Diam{{\sf Diam}}

\newcommand{\nb}[1]{}

\begin{document}
\title[Optimal Distributed Edge-Biconnectivity]{An Optimal Distributed Edge-Biconnectivity Algorithm}
\author[D. Pritchard]{David Pritchard}
\address{Department of Combinatorics and Optimization, Waterloo,
Canada.} \thanks{The author thanks his Master's advisor Santosh
Vempala for his support and for helpful comments on this paper.}
\begin{abstract} We describe a synchronous distributed algorithm which
identifies the edge-biconnected components of a connected network.
It requires a leader, and uses messages of size $O(\log |V|).$ The
main idea is to preorder a BFS spanning tree, and then to
efficiently compute least common ancestors so as to mark cycle
edges. This algorithm takes $O(\Diam)$ time and uses $O(|E|)$
messages. Furthermore, we show that no correct singly-initiated
edge-biconnectivity algorithm can beat either bound on \emph{any}
graph by more than a constant factor. We also describe a
near-optimal local algorithm for edge-biconnectivity.
 \end{abstract}
 \maketitle

\comment{
\begin{enumerate}
\item
 Contact information:\\
 \begin{tabular}{rl}
 Name & David Pritchard \\
 Phone & 617-921-7167 \\
 Fax & 519-725-5441 \\
 Email & daveagp@gmail.com \\
 Postal Address & Combinatorics \& Optimization \\
 & University of Waterloo \\
 & 200 University Avenue West \\
 & Waterloo, Ontario, Canada \\
 & N2L 3G1
 \end{tabular}
\item Title: An Optimal Distributed Edge-Biconnectivity Algorithm
\item Author list: David Pritchard, University of Waterloo
\item Abstract: \\
We describe a synchronous distributed algorithm which identifies the
edge-biconnected components of a connected network. It requires a
leader, and uses messages of size $O(\log |V|).$ The main idea is to
preorder a BFS spanning tree, and then to efficiently compute least
common ancestors so as to mark cycle edges. This algorithm takes
$O(\Diam)$ time and uses $O(|E|)$ messages. Furthermore, we show
that no correct singly-initiated edge-biconnectivity algorithm can
beat either bound on \emph{any} graph by more than a constant
factor. We also describe a near-optimal local algorithm for
edge-biconnectivity.
\item Keywords: edge-connectivity, biconnected components,
bridge, universally optimal, spanning trees, preordering, least
common ancestor, local algorithm, will-maintaining algorithm
\item Number of Pages: 20 (including this one)
\item {\bf This paper is eligible for the best student paper award.}
\end{enumerate}}

\section{Introduction}
The \emph{edge-biconnectivity} problem is to partition the vertices
of a graph into maximal subsets called \emph{components} such that
the subgraph induced by each component remains connected after the
deletion of any one edge. An edge whose deletion disconnects a graph
is called a \emph{bridge}; identification of all bridges in a graph
is roughly equivalent to computing its components.

Here is a simple application of edge-biconnectivity. Given a
(connected) communication network, let us compute its components.
Two members of the network would be able to communicate despite the
failure of any one communication link if and only if they are in the
same component. Furthermore, for a given link, some members need
that link to communicate if and only if that link is a bridge. This
problem also has more sophisticated applications to module
dependency \cite{incremental-bcc} and efficient fault-tolerant
broadcast\cite{itairodeh}.

Connectivity is also important as a tool in graph theory. For
example, vertex connectivity plays a major role in the theories of
excluded minors and embeddings \cite{lovasz-minor}\cite{papa-conj}.
There are sequential algorithms for $k$-connectivity that are
optimally efficient
--- having $O(|V|+|E|)$ time complexity --- when $k$ is small
\cite{tarjandfs}\cite{tri-tarjan}\cite{tarjan74}. In this paper we
give an optimal edge-biconnectivity $(k=2)$ algorithm for
distributed networks. Although this has been claimed before
\cite{ghetto}, our algorithm is optimal in a stronger sense: no
correct distributed algorithm can outperform it on \emph{any} graph.

A recurring point will be that the ``optimality" of an algorithm
depends on what preliminary assumptions are made. This seems to come
up in distributed algorithms more than in sequential algorithms due
to the variety of models. Our claims assume a synchronous network
with a leader, and that the algorithm is event-driven, starting with
a single initiator. These are the same assumptions made by the
previous best edge-biconnectivity algorithm \cite{thur}. However, we
also discuss how our algorithm performs under different assumptions.

\section{Preliminaries} \label{prelim}

\comment{ Let $\G$ be a connected unweighted graph. An edge of $\G$
is said to be a \emph{bridge} if its removal causes $\G$ to become
disconnected. The \emph{edge-biconnected components} of $\G$ are the
connected components of $\G$ after we delete all bridges.
Equivalently, the edge-biconnected components are the maximal
induced subgraphs of $\G$ which remain connected even after an edge
is deleted. The \emph{edge-biconnectivity} problem is to identify
all bridges and edge-biconnected components of $\G.$ One application
is that, in designing a network, the bridges are the points which
make the network most vulnerable to failure or attack. Where there
is no confusion, we simply write ``biconnectivity" instead of
``edge-biconnectivity."}

Throughout this paper, we write $\G = (V, E)$ for a connected graph,
with $n=|V|$ and $m=|E|.$ We write $\Diam$ for the maximum distance
between any two vertices of $\G.$ We take $\G$ to model a computer
network, with nodes representing computers and edges representing
two-way, reliable, communication links.

Let us precisely state what edge-biconnectivity means. Define a
relation $\sim$ on $V(\G)$ by $x \sim y$ if, despite the removal of
any one edge from $\G,$ there remains a path from $x$ to $y.$ It is
easy to show that this is an equivalence relation, and we define the
edge-biconnected components to be its equivalence classes. An edge
is a bridge if its deletion causes $\G$ to become disconnected. The
connection between components, bridges, and cycles is shown by the
following lemma, whose proof we postpone until Section \ref{bd}.
\begin{lemma} \label{lemma2} For a connected graph $\G$ and $(x, y) \in E(\G),$
the following are equivalent:
\begin{enumerate}
\item $x$ and $y$ are in the same biconnected component,
\item $(x, y)$ is not a bridge,
\item $x$ and $y$ lie in a simple cycle.
\end{enumerate}
\end{lemma}

We will give a distributed algorithm for computing the
edge-biconnected components of any graph. Where there is no
confusion, we write ``biconnectivity" instead of
``edge-biconnectivity." At the end of the algorithm, each node will
store a label corresponding to its biconnected component. Thus, by
Lemma \ref{lemma2}, a given edge is a bridge if and only if its two
endpoints store different labels.

We will assume that the network is synchronous. This common
assumption is validated by synchronizers \cite{synch85}, by which
our synchronous algorithm can be efficiently made into an
asynchronous algorithm. We also assume that the network initially
contains a distinguished node called the \emph{leader}. If we do not
make this assumption, then we instead use an existing leader
election algorithm \cite{awer87} at an overhead of $O(n)$ time and
$O(m + n \log n)$ messages.

This biconnectivity algorithm communicates, for the most part, along
a spanning tree of the network. The key point is that we can use a
tree to efficiently identify which edges lie in cycles, and then
apply Lemma \ref{lemma2}. We distributively compute a pre-ordering
of $V,$ and using ``least common ancestors" we efficiently identify
all cycle edges. Then it is straightforward to identify the bridges.
Finally, we label the components using the following lemma.
\begin{lemma} \label{lemma3} If we delete all bridges of $\G$ from a spanning tree of $\G,$
 then the forest of
resulting trees is a set of spanning trees for the biconnected
components of $\G.$
\end{lemma}

The important point is that, in comparison to DFS-based distributed
biconnectivity algorithms such as \cite{ahujazhu} \cite{hohberg}
\cite{ghetto}, our algorithm works on \emph{any} tree. Whereas DFS
seems to require $\Omega(n)$ time, using BFS in our algorithm keeps
the time complexity low. The idea of using an arbitrary tree for
bridge-finding was first published in a 1974 paper by
Tarjan\cite{tarjan74}, and was generalized somewhat in
\cite{generalbridge}.

All messages in the algorithm are $O(\log n)$ bits long, and thus it
meets the $\mathcal{CONGEST}$ model of \cite{p2000}.  Our algorithm
takes $\Theta(\Diam)$ time and sends $\Theta(m)$ messages. The
previous best time complexity for a biconnectivity algorithm is
$O(\Diam + n^{0.614})$ from \cite{thur}. Given that this is the same
as our bound when $\Diam > n^{0.614},$ one may wonder whether we
have really improved the situation. Happily, we can show that this
new algorithm is \emph{universally optimal:} any event-driven
biconnectivity algorithm with a single initiator must take
$\Omega(\Diam(\G))$ time and use $\Omega(m(\G))$ messages for
\emph{every} possible graph $\G,$ or else the algorithm is incorrect
on some graphs.

\comment{Towards the end, we discuss some extensions of our
algorithm. For one, we do away with the leader, and instead start
all vertices simultaneously. We define a parameter which we call the
\emph{cycle-witness radius.} This parameter serves as a lower bound
on the time complexity of a globally-initiated biconnectivity
algorithm. Furthermore, using neighbourhood covers, we give a
will-maintaining algorithm which exceeds this time bound by only
logarithmic factors. We also discuss other types of connectivity:
triconnectivity, strong connectivity, and vertex connectivity.}

For reference, we note other previous work on biconnectivity. The
distributed algorithm of \cite{huang} is the most similar to ours,
as it takes an arbitrary tree, but it uses $O(mn)$ messages. The
parallel algorithms of \cite{tarjan-parallel}\cite{hyper-bicon} also
use an arbitrary tree. The distributed algorithm of \cite{eardecomp}
uses an ``ear decomposition" of the network and takes $O(n)$ time.
Several others \cite{ghetto}\cite{chang}\cite{inria} use messages of
size $\Omega(n)$ bits, including an incremental algorithm
\cite{incremental-bcc} and a self-stabilizing algorithm
\cite{karaata}. Another self-stabilizing algorithm \cite{chaudhuri2}
takes $O(n^2)$ time.

\section{Overview of the Algorithm}
We use a rooted tree $\tree$ in our algorithm. An edge of $\G$ which
does not lie in $\tree$ is called a \emph{cross edge.} Let
$h(\tree)$ denote the height of $\tree$, and $\Desc(v)$ denote the
descendants of $v$ in $\tree$, including $v$ itself. Let $\cyc$
denote the union of all simple cycles, $$\cyc := \{e \in E \mid  e
\textrm{ lies within some simple cycle of } \G\}.$$

The algorithm operates in five phases, as follows:
\begin{enumerate}
\item Construct a rooted breadth-first search (BFS) tree $\mathcal{T}.$
\item At each node $v$, compute $\NDesc(v)$, the number of descendants of $v$ in $\mathcal{T}$.
\item Compute a preorder labeling of $V(\G)$ with respect to $\tree$.
\item By sending messages from cross edges up to the root, mark each edge in
$\mathcal{C}$. \comment{During this phase we identify the bridges
and articulation points, and identify generators for $\sim.$ }
\item By downcasting, label the nodes
according to their biconnected components.
\end{enumerate}

\section{Tree Construction and Preorder Labeling}
First, we need a rooted spanning tree $\tree$. The algorithm runs
fastest when $\tree$ is a BFS tree, but for the purposes of
correctness any tree will do. We assumed that there is a leader in
the network, and given this leader, it is straightforward to
construct a rooted spanning tree. This is a well-studied problem,
see for example \cite{p2000}. In what follows, we use the term
\emph{downcasting} to mean that the root sends a message to all of
its children, each of which sends a message to each of its children,
and so forth. \emph{Convergecasting} \cite{p2000} means an inverse
process, where messages are propagated from leaves to their parents,
and so on up the tree to the root; however, each node waits to hear
from all of its children before reporting to its parent, so only
$n-1$ messages are sent in total.

The computation of $\NDesc(v)$ at each node in Phase 2 can be
accomplished in 2$h(\tree)$ time steps. First, the root node sends
``Compute $\NDesc$ of yourself" to each of its children, and this
message is downcasted through all of $\mathcal{T}$. Each leaf $v$
determines immediately that $\NDesc(v)=1.$ Once any non-root node
computes its $\NDesc$ value, it sends a message to its parent
indicating that value. Each non-leaf node $v$ aggregates its
childrens' values in order to compute $\NDesc(v);$ namely, if $v$
has children $c_1, \dotsc, c_k,$ then $v$ may compute
$$\NDesc(v) := 1 + \sum_{i=1}^k \NDesc(c_i).$$

In Phase 3, a preorder labeling of $\tree$ is computed by using
another downcast. The root node starts by setting its own $PreLabel$
field to 1. Whenever a node $v$ sets its $PreLabel$ field to $\ell$,
it orders its children in $\mathcal{T}$ arbitrarily as $c_1, c_2,
\ldots$. Then $v$ sends the message ``Set your $PreLabel$ field to
$\ell_i$" to each $c_i$, where $\ell_i$ is computed by $v$ as
$$\ell_i  = \ell+1+\sum_{j < i} \NDesc(c_j)$$
After $h(\mathcal{T})$ time steps, we will have computed a
preordering of $\mathcal{T}$.

\section{Least Common Ancestors} In order to simplify the
presentation, we hereafter refer to nodes simply by their preorder
labels. The preordering allows us to reduce congestion in Phase 4 of
the algorithm, using the following properties.
\begin{lemma} \label{lemma:desc}
The descendants of a node $v$ in the tree \tree are precisely
$$\Desc(v) = \{ u \mid v \leq u < v + \NDesc(v)\}.$$
\end{lemma}
Let $\LCA(u_1, u_2, \ldots)$ denote the lowest (by position, not
value) common ancestor of nodes $u_1, u_2, \ldots$ in the tree
\tree. In other words, $\LCA(u_1, u_2, \ldots)$ is an ancestor of
each $u_i,$ but no strict descendant of $\LCA(u_1, u_2, \ldots)$ is
an ancestor of all $u_i$s.

\begin{theorem}
If $v_1 \leq v_2 \leq v_3$, then $\LCA(v_1, v_3)$ is an ancestor of
$v_2$.
\end{theorem}
\begin{proof}
Let $a = \LCA(v_1, v_3)$. By Lemma \ref{lemma:desc}, $a \leq v_1
\leq v_3 < a + \NDesc(a)$. Thus $a\leq v_2 < a + \NDesc(a)$, and by
Lemma \ref{lemma:desc}, $v_2$ must also be a descendant of $a$.
\end{proof}
\begin{corollary}
$\LCA(u_1, u_2, \ldots u_k) = \LCA(\min_i(u_i), \max_i(u_i)).$
\end{corollary}
\begin{corollary} \label{cor2}
If $u_i \leq v_i$ for all $i$, then $$\LCA(\LCA(u_1, v_1), \LCA(u_2,
v_2), \ldots \LCA(u_k, v_k)) = \LCA(\min_i(u_i), \max_i(v_i)).$$
\end{corollary}

\section{Marking Cycle Edges} The goal of Phase 4 is to determine
which edges lie in $\mathcal{C}.$  When $v'$ is an ancestor of $v$
in \tree, let $\Chain(v', v)$ denote the set of edges on the path
from $v'$ to $v$ in \tree. The cross edges with respect to $\tree$
permit a simple formula for $\cyc$:
 \begin{lemma} \label{lemma:eq}
\begin{equation}\label{eq:union}
\cyc = \bigcup_{(u, v) \in \G - \tree} \{(u, v)\} \cup
\Chain(\LCA(u, v), u) \cup \Chain(\LCA(u, v), v).\end{equation}
\end{lemma}
\begin{proof}
Note that each set $\{(u, v)\} \cup \Chain(\LCA(u, v), u) \cup
\Chain(\LCA(u, v), v)$ is a simple cycle. It remains to show that
this union formula contains \emph{all} edges appearing in simple
cycles. Suppose otherwise, that the above formula missed some edge
$(u, v)$ belonging to a simple cycle $K$ of $\G$. Since Equation
\eqref{eq:union} includes all edges of $\G - \tree,$ we can assume
that $(u, v) \in \tree$, without loss of generality $u$ the parent
of $v$.

Let the cycle $K$ contain, in order, the nodes $(k_0=v, k_1, k_2,
\ldots, k_{m-1} = u, k_m = v)$. If $k_i$ is the first element of
this list not in $\Desc(v)$, then $(k_{i-1}, k_i)$ is a cross edge.
But then we have
\begin{equation*}
(u, v) \in \Chain(\LCA(k_{i-1}, k_i), k_{i-1}),
\end{equation*}
so in fact $(u, v)$ is counted by \eqref{eq:union}.
\end{proof}
Thus, to mark the edges of $\cyc$, it suffices to just mark chains
going up from each cross edge to its endpoints' LCA.

We could distributively mark the edges in $\Chain(\LCA(u, v), v)$ as
follows:
\begin{itemize}
\item For each cross edge $(u, v)$,
\subitem Send a message from $v$ to $u$ which states ``If you are an
ancestor of both $u$ and $v,$ then ignore this message. Otherwise,
pass this message up to your parent, and mark the edge joining you
to your parent as being in \cyc." \subitem Send the same message
from $u$ to $v$.
\end{itemize}
Checking the ancestry condition is accomplished using Lemma
\ref{lemma:desc}. We will abbreviate the message ``If you are an
ancestor of both $u$ and $v,$ \ldots" as ``Mark up to $\LCA(u, v)$".
Without loss of generality, we will send our messages so that $u
\leq v$.

Sending these messages na\"ively leads to congestion. When a node
receives many at once that must all be forwarded, not all can be
immediately sent to its parent if the $O(\log n)$ bound on message
sizes is to be respected. The following Forwarding Rule fixes this
congestion:
\begin{itemize}
\item If a node $w$ receives several messages ``Mark up to
$\LCA(u_i, v_i)$" for $i=1 \ldots k$, it should compute $u_{\min} =
\min_i{u_i}$ and $v_{\max} = \max_i{v_i}$. If $w$ is an ancestor of
both $u_{\min}$ and $v_{\max}$, then no message is sent up.
Otherwise, $w$ should send ``Mark up to $\LCA(u_{\min}, v_{\max})$"
to its parent, and mark the edge connecting $w$ to its parent as
being in \cyc.
\end{itemize}
\begin{theorem} \label{thm1}
The Forwarding Rule correctly marks $\Chain(\LCA(u, v), v)$ for each
cross edge $(u, v)$.
\end{theorem}
\begin{proof}
Suppose that $w$, as described, is asked to propagate messages so
that all edges in $$\bigcup_i \Chain(\LCA(u_i, v_i), w)$$ become
marked. They must all lie on the unique path between $w$ and the
root of \tree, so we only need to mark the longest chain. The
highest LCA is equal to $\LCA(\LCA(u_1, v_1), \LCA(u_2, v_2),
\ldots)$ and by Corollary \ref{cor2} this is $\LCA(u_{\min},
v_{\max})$, so the propagated message (if any) is correct.
\end{proof}

\section{Biconnected Decomposition} \label{bd}
\begin{lemma} For a connected graph $\G$ and $x, y \in V(\G),$
$x$ and $y$ are in the same biconnected component if and only if $x$
and $y$ both lie in some cycle of $\G$ that has no repeated edges.
\label{lemma1}
\end{lemma}
\begin{proof} Note that $x \not\sim y$
if and only some edge's deletion separates $x$ from $y;$ by the
maxflow-mincut theorem such an edge exists if and only if there are
not two edge-disjoint paths between $x$ and $y.$ Those paths' union
is precisely a cycle with no repeated edges, and likewise such a
cycle can be broken into two edge-disjoint $x$-$y$ paths.
\end{proof}

We now prove the Lemma introduced in Section \ref{prelim}.

{\noindent \bf Lemma \ref{lemma2}.} \emph{For a connected graph $\G$
and $(x, y) \in E(\G),$ the following are equivalent:
\begin{enumerate}
\item $x$ and $y$ are in the same biconnected component,
\item $(x, y)$ is not a bridge,
\item $x$ and $y$ lie in a simple cycle.
\end{enumerate}
}
\begin{proof} Let $e$ denote the edge
$(x, y).$

$(3)\Rightarrow(2):$ Let $C$ be a cycle containing $x$ and $y.$ For
any $u$-$v$ path containing $e,$ we can use $C - e$ to connect $u$
and $v$ in $\G - e,$ and thus $e$ is not a bridge.

$\neg(1)\Rightarrow\neg(2):$ If $x \not\sim y,$ then the deletion of
some edge from $\G$ causes $x$ and $y$ to become separated. But $x$
remains connected to $y$ by $e$ unless $e$ is the deleted edge. Thus
the deletion of $e$ from $\G$ separates $x$ from $y,$ so $e$ is a
bridge.

$(1)\Rightarrow(3):$ Apply Lemma \ref{lemma1}, obtaining cycle $C$
with no repeated edges. If $e \not\in C,$ then we take any simple
$x$-$y$ path in $C$ and adjoin $e$ to create a simple cycle
containing $x$ and $y.$ If $e \in C,$ then partition $C$ into simple
cycles, and take the one containing $e.$
\end{proof}

Note that only tree edges can be bridges, for each cross-edge
induces a cycle with $\tree.$ Thus, if each node stores a boolean
variable indicating whether the edge to its parent is a bridge, then
this suffices to identify all bridges.

To group the nodes according to their biconnected component (Phase
5), we need to broadcast an identifier along each component. The
following claim means that a simple downcast along the edges of
$\tree$ will suffice.

{\noindent \bf Lemma \ref{lemma3}.} \emph{If we delete all bridges
of $\G$ from $\tree,$ then the forest of resulting trees is a set of
spanning trees for the biconnected components of $\G.$}
\begin{proof}
Suppose otherwise, that there are two nodes $u \sim v$ such that the
unique path $P$ between $u$ and $v$ in $\tree$ contains a bridge
$e.$ Since $u \sim v,$ there is a simple $u$-$v$ path $P'$ in $\G -
e.$ But then the symmetric difference $P \triangle P'$ has even
degree at every node and contains $e,$ and so $P \triangle P'$
contains some simple cycle containing $e$. By Lemma \ref{lemma2} we
have a contradiction.
\end{proof}

\section{Correctness and Complexity of the Algorithm}
The main parts of the algorithm, Phases 4 and 5, are shown in
Algorithm \ref{alg:bcc}. Its correctness follows from Lemma
\ref{lemma:eq}, Theorem \ref{thm1} and Lemma \ref{lemma3}. Note that
we specify ``null'' messages which we have not yet been justified;
this is in order for the ``Mark up to'' messages to be properly
synchronized in a convergecast. This way, a node simply waits to
hear from all of its non-parent neighbours before reporting to its
parent.

\begin{figure} \caption{Distributed algorithm for
edge-biconnectivity, given a rooted spanning tree.} \label{alg:bcc}
[Algorithm specification omitted, as arxiv does not support the
algpseudocode class].
\end{figure}

Phases 2--4 have total message complexity $O(n)$ and time complexity
$O(h(\tree)),$ even in an asynchronous setting. Thus, those phases
would be optimized when $\tree$ is as short as possible. Note that a
BFS tree has height at most $\Diam(\G),$ and that no spanning tree
has height less than $\Diam(\G)/2.$ Thus, it is essentially optimal
to construct a BFS tree in Phase 1. Assuming synchrony, and that we
are given a leader at the beginning of the algorithm, it is
well-known that a BFS tree can be constructed greedily in
$\Theta(\Diam)$ time and using $O(m)$ messages. Thus, the
biconnectivity algorithm's total complexity is $O(\Diam)$ time and
$O(m)$ messages.

\section{Optimality} \label{sec:optim}
We claim that the performance of this algorithm cannot be improved
beyond constant factors. To be precise, we argue that any
deterministic, singly-initiated, event-driven protocol for
bridge-finding must always send at least $m$ messages and take at
least $\Diam/2$ time, or else the protocol will not work on all
graphs. By singly-initiated, we mean that there is a single node in
the graph which begins computing spontaneously, and by event-driven
we mean that every other node must receive a message before it can
perform any action. These lower bounds are similar in nature, and
both depend on the fact that the whole network must be explored.

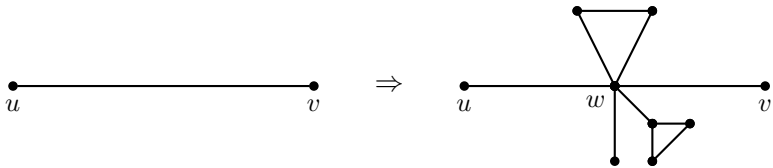
\begin{figure}
\begin{center}
\begin{pspicture}*(-9, -1.5)(3,1.5)
    \uput[270](-8,0){$u$}
    \uput[270](-4,0){$v$}
    \pcline{*-*}(-8,0)(-4,0)
    \rput(-3,0){{\large $\Rightarrow$}}
    \uput[270](-2,0){$u$}
    \uput[225](0,0){$w$}
    \uput[270](2,0){$v$}
    \pcline{*-*}(-2,0)(0,0)
    \pcline{*-*}(2,0)(0,0)
    \pcline{*-*}(0,0)(0,-1)
    \pcline{*-*}(0,0)(-0.5,1)
    \pcline{*-*}(0,0)(0.5,1)
    \pcline{*-*}(-0.5,1)(0.5,1)
    \pcline{*-*}(0,0)(0.5,-0.5)
    \pcline{*-*}(0.5,-1)(0.5,-0.5)
    \pcline{*-*}(1,-0.5)(0.5,-0.5)
    \pcline{*-*}(1,-0.5)(0.5,-1)
\end{pspicture}
\end{center}
\caption{Modification of $\G$ into $\G',$ upon which a
biconnectivity algorithm fails.} \label{fig:edge}
\end{figure}

First we describe the lower bound for messages. Suppose that, when
the protocol is executed on some graph $\G,$ there is an edge
$(u,v)$ along which no messages are sent. Let $\G'$ be a graph
obtained from $\G$ by adding a new node $w$ and dividing $(u,v)$
into two edges $(u,w)$ and $(w,v).$ We also attach some cycles and
bridges to $w,$ as shown in Figure \ref{fig:edge}. When we run the
protocol on $\G',$ assuming that the algorithm is deterministic, no
messages are sent along $(u,w)$ or $(w,v)$ and so no messages reach
the new nodes and edges. Consequently, the algorithm cannot
correctly determine whether the new edges are bridges or not.

The lower bound on the time complexity is similar. If an algorithm
uses takes less than $\Diam(\G)/2$ steps on some graph $\G,$ then
there are parts of the graph which no messages reach. Consequently,
we can modify $\G$ so that the algorithm operates incorrectly.
\comment{Similarly, these lower bounds of $\Omega(m)$ messages and
$\Omega(\Diam(G))$ time hold for any problem which necessarily
involves some computation in every small neighbourhood of $\G.$}

Note that these lower bounds apply to \emph{all} graphs. In
comparison, a $O(n)$-time algorithm of \cite{ghetto} was called
``optimal" because \emph{some} graphs require $O(n)$ time to find
their bridges. The $O(n)$ algorithm is \emph{existentially optimal,}
 since there exist some instances on which the protocol is optimal, and
our $O(\Diam)$ algorithm is \emph{universally optimal,} since it has
optimal running time on all instances. The different types of
optimality were first observed by \cite{pelegopt} and \cite{awerbfs}
in the context of leader election, and further discussion appears in
\cite{sublinear-mst} and \cite{cycrad}. Universal optimality allows
us to precisely state that the inherent complexity of the
(singly-initiated, event-driven) biconnectivity problem is
$\Theta(\Diam)$ time and $\Theta(m)$ messages. Finally, although we
state these bounds for deterministic algorithms, similar bounds can
be proved for randomized ones, using essentially the same argument.

\section{A Near-Optimal Local Algorithm} \label{sec:sublinear-bcc}
For now, let us forget the problem of labeling nodes according to
their biconnected component, and only worry about identifying all of
the bridges in a graph. We consider initiating all nodes at the same
time, and want to know how long it will be before all edges are
correctly identified as bridges or non-bridges. By removing the
assumption of a single initiator, we can beat the lower time bound
of $\Diam$.

Suppose we remove the restriction on the message size. Then, each
node can broadcast everything it knows about its local topology
after each step, and so after $t$ steps each node will know its own
$t$-neighbourhood. Here is an algorithm for bridge-finding.
Initially, each edge is assumed to be a bridge; whenever a node
learns of a cycle in its neighbourhood, it informs all of the edges
in that cycle that they are not bridges. In this way we
distributively determine $\cyc,$ the union of all cycles in $\G.$
For a general graph, we cannot be sure that we're done until $\Diam$
steps have elapsed. However, for certain graphs, we can safely
terminate in $o(\Diam)$ rounds. An algorithm whose time complexity
is $o(\Diam)$ is often called \emph{local}.

Now, let us determine the time before this algorithm has correctly
identified the non-bridges. A non-bridge $e$ will be identified as
soon as a cycle containing $e$ is known by a node; we call such a
cycle a \emph{witness} for $e.$ We need each edge to be identified
by a witness in order for the algorithm to be correct. Define the
\emph{cycle-witness radius} of $\G,$ denoted $\Upsilon(\G),$ by
$$\Upsilon(\G) := \max_{e \in \cyc} \min_{\substack{K \textrm{ a
cycle } \\ K \ni e}} \min_{v \in V(\G)} \max_{u \in K} dist_\G(u,
v).$$ Then the cycle-witness radius is the minimum time needed to
identify all of the non-bridges (and, it will take another
$\Upsilon(\G)$ rounds to notify those edges). Further, it can be
shown that $\Upsilon(\G)$ is a lower bound on the number of rounds
before all non-bridges can be correctly identified.

A $(\log n, \Upsilon)$-neighbourhood cover of $\G$ is a collection
of connected vertex sets called \emph{clusters} such that
\begin{enumerate}
\item For each vertex $v,$ the $\Upsilon$-neighborhood
of $v$ is entirely contained in some cluster.
\item The subgraph of $\G$ induced by each cluster has diameter $O(\Upsilon \log n ).$
\item Each node belongs to $O(\log n)$ clusters.
\end{enumerate}
See \cite[Ch.\ 21]{p2000} for a good exposition on this subject. The
full version of \cite{cycrad} gives a randomized \emph{local}
algorithm for computing sparse neighborhood covers which, with high
probability, runs in $O(\Upsilon \log^3 n)$ time and uses $O(m
\log^2 n)$ messages on a synchronous network.

We can use neighbourhood covers to search brute-force for all small
cycles, without using large messages as described above. Suppose we
have a $(\log n, \Upsilon)$ neighbourhood cover as described above.
Each edge has a witness that is entirely contained within one
cluster. If we run Algorithm \ref{alg:bcc} separately on each
cluster, each non-bridge will be witnessed in \emph{some} cluster.
Since each node may be in $O(\log n)$ clusters, there will be
congestion when we process all clusters in parallel; however, this
will only increase the time of the biconnectivity algorithm by a
factor of $O(\log n),$ since each node can rotate between
participating in its containing clusters. The resulting local
algorithm takes $O(\Upsilon \log^3 n)$ time and $O(m \log^2 n)$
messages to construct the clusters, then a further $O(\Upsilon \log
n)$ time and $O(m \log n)$ messages to determine the non-bridges.

Finally, it is unlikely that $\Upsilon$ can be computed efficiently
and/or locally. However, an algorithm can successively ``guess"
$\Upsilon=1, 2, 4, 8, \dotsc,$ and run the local algorithm for each
value in turn. Once the guess is larger than the actual value of
$\Upsilon,$ all edges will be correctly classified; this algorithm
becomes correct within $O(\Upsilon \log^3 n)$ rounds. We note that
this is essentially a \emph{will-maintaining algorithm} as defined
by Elkin in \cite{cycrad}.

\section{Other Extensions} \label{sec:scc}
With a small modification, the algorithm of this chapter can also be
used to compute the strongly-connected components of a graph. We
require that all directed edges function as 2-way communication
channels. We compute a directed DFS tree $\tree$ of the network. It
is easy to show that an analog of Equation \eqref{eq:union} holds in
this case, with the cross edges replaced by the back edges $\{(u, v)
\mid u \in desc(v)\}.$ The resulting algorithm takes $O(h(\tree))$
time and $O(m)$ messages, identifies the edges that belong to
cycles, and labels all nodes according to their strongly-connected
component. Using Awerbuch's DFS algorithm from \cite{awerdfs} gives
a total of $O(n)$ time and $O(m)$ communication complexity, and note
that it works on asynchronous networks.

If a short DFS tree could be identified in sub-linear time, then we
might be able to get a sub-linear algorithm for identifying strongly
connected components.
\begin{question} Does there exist a $(O(\Diam)+o(n))$-time distributed DFS tree construction
algorithm, using messages of size $O(\log n)$?
\end{question}
Also, there is a divide-and-conquer algorithm \cite{dc-scc} for
strongly connected components which might lend itself to distributed
implementation in sub-linear time.

In order to ensure reliability in networks, one may want to ensure
the 2-\emph{vertex} connectivity of a network, for example
\cite{itairodeh}. It does not seem that our algorithm can be easily
modified to determine vertex biconnectivity. In contrast, the
$\Theta(n)$ time DFS-based biconnectivity algorithms of
\cite{ahujazhu} and \cite{hohberg} can determine vertex
biconnectivity.

We might also try to determine the \emph{triconnected}
\cite{tri-tarjan} components of a graph. There are efficient
parallel algorithms for this problem \cite{tri-replace}. The
following lemma might be useful in designing a fast distributed
triconnectivity algorithm.
\begin{claim} \label{claim:tricon}
Let $\G$ be a graph with no bridges. Define the relation $\sim_D$ on
the edges of $\G$  by $x \sim_D y$ if the graph $\G - x - y$ is not
connected. Then $\sim_D$ is an equivalence relation.
\end{claim}

\newcommand{\noopsort}[1]{} \newcommand{\printfirst}[2]{#1}
  \newcommand{\singleletter}[1]{#1} \newcommand{\switchargs}[2]{#2#1}

\end{document}